# Intrinsic pH of water/vapor interface revealed by ion-induced water alignment

Kuo-Yang Chiang, Laetitia Dalstein, and Yu-Chieh Wen[*]

Institute of Physics, Academia Sinica, Taipei 11529, Taiwan, R. O. C.

*Correspondence to: ycwen@phys.sinica.edu.tw (Y.C.W.)

**ABTRACT:**

Protons at the water/vapor interface are relevant for atmospheric and environmental processes, yet to characterize their surface affinity on the quantitative level is still challenging. Here we utilize phase-sensitive sum-frequency vibrational spectroscopy to quantify the surface density of protons (or their hydronium form) at the intrinsic water/vapor interface, through inspecting the surface-field-induced alignment of water molecules in the electrical double layer of ions. With hydrogen halides in water, the surface adsorption of protons is found to be independent of specific proton-halide anion interactions and to follow a constant adsorption free energy, $\Delta G \sim -3.74\ (\pm 0.56)$ kJ/mol, corresponding to a reduction of the surface pH with respect to the bulk value by 0.66 ($\pm 0.10$), for bulk ion concentrations up to 0.3 M. Our spectroscopic study is not only of importance in atmospheric chemistry, but also offers a microscopic-level basis to develop advanced quantum-mechanical models for molecular simulations.

**MAIN TEXT:**

Protons at water/vapor interfaces play a key role in atmospheric chemistry and environmental science (1, 2). They have been the topics of extensive theoretical and experimental studies for decades (3-18), but good understanding of the surface affinity of protons on the microscopic and quantitative level is still lacking. Molecular dynamics (MD) simulations generally predicted that protons ($H^+$) could appear at the interface in the form of hydronium ($H_3O^+$), while the quantitative details, such as the adsorption free energy and the depth profile of ion concentrations, depend very much on the molecular model and interaction potentials assumed (3-8). Several experimental approaches have been used to verify the qualitative conclusion from the simulations (9-18), but extension of these efforts to quantitative characterizations of the proton adsorption remains challenging (17). This largely limits our knowledge on the underlying mechanisms and the role of proton in many relevant interfacial processes. Technically, it is generally difficult to adopt macroscopic observables from surface tension, electrokinetic, and Kelvin probe measurements to yield molecular-level interfacial properties without crucial model assumption (17-19). For molecular-scale measurements, X-ray spectroscopies can examine the proton effect at water interfaces indirectly through probing relative concentration changes of the counterions (9, 10), but, yet, the surface affinity of protons has not been quantified from the results. Surface-specific optical sum frequency generation (SFG) and second harmonic generation (SHG) were widely employed to explore this issue (11-16). But, in many reports, the proton propensity were only studied *comparatively* with respect to other ions without quantitative details (13-15). In the other, the surface concentration of protons was estimated by inspecting their reaction with *extrinsic* interfacial label molecules (16), whereas how the result is affected by the proton-label molecule binding interaction remains an open question.

The general difficulty of the SFG/SHG spectroscopy in quantifying charged species at water interfaces stems from our long-term inability to separate two signal contributions (20-29): One comes from an interfacial molecular layer where the inversion symmetry is broken by the interface-specific bonding structure. [This region is labelled as "the bonded interface layer (BIL)" (22, 24).] The other is induced by water molecules in the electrical double layer (EDL) of ions with their inversion symmetry broken by the dc electric field (mainly through the field-induced molecular alignment) (20-23). In recent years, advances in the SFG/SHG spectroscopy now allow unique spectral separation of the BIL and EDL (24-27). With the EDL signal being a probe for the dc field, one can determine the absolute surface charge (ion) density as the third-order nonlinear susceptibility, $\overleftrightarrow{\chi}_B^{(3)}$, that characterizes the response of bulk water to the dc and optical fields is known (26).

In this paper, we report the first spectroscopic determination of the surface proton density, i.e., surface pH, of the intrinsic water/vapor interface. We utilize the phase-sensitive sum-frequency vibrational spectroscopy (PS-SFVS) to examine the ion effect on the vibrational spectrum of the BIL, $\overleftrightarrow{\chi}_{BIL}^{(2)}(\omega_{IR})$, and the spectrum of the EDL, $\overleftrightarrow{\chi}_{EDL}^{(2)}(\omega_{IR})$, separately for the acidic-solution surfaces, and quantify the net surface charge density, $\sigma$, from $\overleftrightarrow{\chi}_{EDL}^{(2)}(\omega_{IR})$ with the $\overleftrightarrow{\chi}_B^{(3)}(\omega_{IR})$ characterized previously. Analyses of $\sigma$ versus bulk proton and halide (the counterion) concentrations yield their partitioning coefficients between the surface and bulk water, and indicate an insignificant effect of direct ion-ion interactions on the results.

**Experiment and Theory**

The samples studied were the surfaces of hydrogen halide solutions probed by a broadband PS-SFVS setup (24, 29) with S-, S-, and P- (SSP-) polarized SF, visible, and IR fields (see

*Materials and Methods* for details). Spectra of the effective surface nonlinear susceptibility, $\chi^{(2)}_{S,eff}(\omega_{IR})$, of the sample were measured with normalization against a *z*-cut quartz reference and the Fresnel coefficients removed. Shown in Fig. 1A are a representative set of the measured $\text{Im}\chi^{(2)}_{S,eff}(\omega_{IR})$ spectra in the OH-stretch region for different HCl concentrations in water. (See *SI Appendix*, Fig. S1 for a complete set of concentration-dependent complex $\chi^{(2)}_{S,eff}(\omega_{IR})$ data for HCl, and Fig. S2 and S3 for HBr and HI, respectively.) Our results are essentially similar with these reported earlier (14, 15, 30), whereas high precision of our measurements allows discernment of the ion-induced spectral changes on the millimolar level.

In analyzing the spectra, we follow Ref. (26) to model the interfacial structure and formulate the SFG process of charged water interfaces. The EDL is set up by an atomically thin layer of the adsorbed ions at the immediate neighborhood of the interface (at $z \approx 0$, with $z$ along the surface normal), characterized by $\sigma$, and other diffusely distributed ions near the interface (31). The interfacial water structure is under the influences of the interfacial bonding interaction in the BIL and the dc field distribution, $E_0(z)$, set up by the EDL. $\chi^{(2)}_{S,eff}$ of such an interface can be expressed as (26)

$$\chi^{(2)}_{S,eff} = \chi^{(2)}_{BIL} + \chi^{(2)}_{EDL}$$
$$\chi^{(2)}_{EDL} \equiv \int_0^\infty [\chi^{(3)}(z') \cdot \hat{z} E_0(z')] e^{i\Delta k_z z'} dz' \cong \chi^{(3)}_B \cdot \hat{z} \Psi \quad (1)$$
$$\text{with } \Psi \equiv \int_0^\infty E_0(z') e^{i\Delta k_z z'} dz',$$

and $\Delta k_z$ the phase mismatch of reflected SFG. $\chi^{(2)}_{BIL}$ and $\chi^{(2)}_{EDL}$ denote the second-order nonlinear susceptibility of the BIL and the effective nonlinear susceptibility of the EDL, respectively. We have used the invariant $\chi_B^{(3)}$ of bulk water to approximate the third-order nonlinear susceptibility of water in the EDL, $\chi^{(3)}(z) \cong \chi_B^{(3)}$ (22-24). In describing the ion effect, we denote $\chi_{S,eff}^{(2)}$ of the interface without ion as $\left[\chi_{S,eff}^{(2)}\right]_{\sigma=0} = \left[\chi_{BIL}^{(2)}\right]_{\sigma=0} (\equiv \chi_{BIL,0}^{(2)})$ and the ion-induced change as $\Delta\chi_{S,eff}^{(2)} = \Delta\chi_{BIL}^{(2)} + \chi_{EDL}^{(2)}$. As explained in details below, we shall extract $\chi_{EDL}^{(2)}(\omega_{IR})$ from the measured $\chi_{S,eff}^{(2)}(\omega_{IR})$, and use the Gouy-Chapman (GC) theory to find $\sigma$ from $\chi_{EDL}^{(2)}(\omega_{IR})$ with the known $\chi_B^{(3)}(\omega_{IR})$ of bulk water (24).

**Results and Discussion**

The neat water/vapor interface is adopted as the reference in our analysis. As shown in Fig. 1A, the $\text{Im}\chi_{S,eff}^{(2)}(\omega_{IR})$ spectrum of the neat water surface shows a negative broad continuum in the 3000-3600 cm$^{-1}$ region and a sharp peak at ~3700 cm$^{-1}$ surrounded by a weak positive hump (3600 ~ 3725 cm$^{-1}$). The broad continuum in 3000 ~ 3600 cm$^{-1}$ arises from stretching of hydrogen (H)-bonded OHs of water molecules with a continuous variation of (or dynamically varying) geometries and bonding strengths. The sharp peak at ~3700 cm$^{-1}$ comes from stretching of the dangling OH of *DAA* water molecules in the topmost surface layer. Here *D* and *A* label donor and acceptor H-bonds, respectively, through which a water molecule connects to neighbors. The weak positive hump around the dangling-OH peak (3600 ~ 3725 cm$^{-1}$) has been explained by symmetric stretching of *DAA* molecules with a weak donor bond (30), or the antisymmetric stretching of the *DDA* molecules (32), in the topmost surface layer. For hydrogen halide solutions (HX with X = Cl, Br, or I), our measurements, similar to the earlier studies (12-15), do not reveal detectable spectral change for [HX] < 1 mM, while the surface ion density that follows its bulk concentration could readily vary by 4 orders of magnitudes as reducing pH from 7 to 3. This indicates that in this concentration range, the surface ion density is negligibly small in terms of $\Delta\chi^{(2)}_{S,eff}$, so that $\chi^{(2)}_{S,eff}(\omega_{IR}) \cong \chi^{(2)}_{BIL,0}(\omega_{IR})$.

Discernible spectral changes appear for the acid concentrations above ~1 mM (mainly in 3000 ~ 3600 cm$^{-1}$, see Fig. 1A). Such a change arises from the structural distortion of the BIL via $\Delta\chi^{(2)}_{BIL}$, and/or emergence of an EDL through $\chi^{(2)}_{EDL}$. To clarify its microscopic origin, we examine the Im$\chi^{(2)}_{S,eff}(\omega_{IR})$ spectra in the 3600–3750 cm$^{-1}$ region, where the spectral features are dominantly contributed from water molecules in the topmost surface layer (as the major part of the BIL) (30, 32), but much less induced by $\chi^{(2)}_{EDL}(\omega_{IR}) \left[= \chi^{(3)}_B(\omega_{IR})\Psi\right]$ because $\chi^{(3)}_B(\omega_{IR})$ appears mainly below 3600 cm$^{-1}$ (22, 24). Therefore, Im$\chi^{(2)}_{S,eff}$ in this spectral region offers a selective probe to the BIL structure (14). Shown in Fig. 1B and 1C are the measured Im$\chi^{(2)}_{S,eff}(\omega_{IR})$ in 3600 ~ 3750 cm$^{-1}$ for 0.1 M HCl, HBr, and HI solutions and 0.8 M HI in comparison with that of the neat water/vapor interface. It is found that the dangling-OH peak and the surrounding positive hump in Im$\chi^{(2)}_{S,eff}(\omega_{IR})$ are consistent for [HX] $\leq$ 0.1 M, but become weaker and stronger, respectively, for 0.8 M HI. A detailed concentration-dependent study indicates invariance of these spectral features for [HCl] < 0.8 M, [HBr] < 0.8 M, or [HI] < 0.2 M (see *SI Appendix*, section 2 for details). It suggests a constant structure of the BIL and, thus, $\left[\chi^{(2)}_{BIL}(\omega_{IR})\right]_\sigma \cong \chi^{(2)}_{BIL,0}(\omega_{IR})$ in the entire OH-stretch range under these ion concentrations. This finding is in agreement with a recent *ab initio* MD simulation (4). Materials that further support this argument are discussed in *SI Appendix*, section 6.

With $\chi^{(2)}_{BIL,0}(\omega_{IR})$ deduced from the neat water surface and $\left[\chi^{(2)}_{BIL}(\omega_{IR})\right]_\sigma \cong \chi^{(2)}_{BIL,0}(\omega_{IR})$ for the particular ranges of the ion concentration, we can extract $\chi^{(2)}_{EDL}(\omega_{IR})$ from the measured

$\chi^{(2)}_{S,eff}(\omega_{IR})$ spectra through Eq. (1) with $\chi^{(2)}_{EDL}(\omega_{IR}) \cong \chi^{(2)}_{S,eff}(\omega_{IR}) - \chi^{(2)}_{BIL,0}(\omega_{IR})$. Shown in Fig. 2 are a representative set of the deduced complex $\chi^{(2)}_{EDL}(\omega_{IR})$ for different HCl concentrations (see *SI Appendix*, Fig. S1 for the complete set of data). It is seen that the $\text{Im}\chi^{(2)}_{EDL}(\omega_{IR})$ spectra exhibit a broad OH-stretch band with its strength increasing negatively for higher [HCl]. This indicates that the water molecules in the EDL tend to reorient with their O→H pointing downward into the bulk water, and thus reflects emergence of a downward dc surface field and the dominant role of cations (hydronium ions) on the surface charging (33). Results for HBr and HI show similar qualitative characteristics, while the strength of the negative OH-stretch band in $\text{Im}\chi^{(2)}_{EDL}(\omega_{IR})$ decreases from HCl, HBr, to HI for a given bulk ion concentration (see *SI Appendix*, Fig. S1–S3 for details). These observations indicate a different dc-field strength and $\sigma$ due to dissimilar surface affinity for different halide species (15, 34).

More quantitatively, we use the GC theory to calculate $E_0(z)$ for a given $\sigma$ and ionic strength (31), and then follow Eq. (1) to calculate $\chi^{(2)}_{EDL}(\omega_{IR})$ from $E_0(z)$ with the known $\Delta k_z$ and the $\chi^{(3)}_B(\omega_{IR})$ spectrum of bulk water (24). Fitting the measured $\chi^{(2)}_{EDL}(\omega_{IR})$ spectra with this calculation yields the value of $\sigma$. As plotted in Fig. 2 (and *SI Appendix*, Fig. S1–S3), the fitting quality for the complex $\chi^{(2)}_{EDL}(\omega_{IR})$ is found to be reasonably well, supporting validity of our analysis. Fig. 3A shows $\sigma$ deduced by fitting the measured $\chi^{(2)}_{EDL}(\omega_{IR})$ spectra for different HX concentrations in water. (High-concentration data that may have assumption-related concerns are depicted by open dots for clarity. Explanations are given below in details.) Again, $\sigma$ is found to be positive and differ for distinct halide species due to the dominant role of cations on $\sigma$ and the ion-specific surface affinity, respectively. Furthermore, $\sigma$ exhibits a clear tendency to increase with [HX] by following a simple linear relation, suggesting that the ion adsorption obeys a constant partitioning coefficient relating surface and bulk ion concentrations. This property that reflects a weak correlation between ions could be explained by the low molar fraction of ions ($10^{-5} \sim 10^{-2}$) in this study (35).

To characterize the partitioning coefficient $K$ for each ionic species from the measured $\sigma$, we consider $H_3O^+$ and $X^-$ to have the surface ion densities $\rho$ and define their $K$ as $K(H_3O^+) \equiv \rho(H_3O^+)/[H^+]$ and $K(X^-) \equiv \rho(X^-)/[X^-]$, but ignore OH$^-$ ion for its much mirror role at acidic pH. So, we have $\sigma = |e|[\rho(H_3O^+) - \rho(X^-)]$ with $|e|$ the elementary charge. With the weak correlation between ions (as found from $\sigma$ versus [HX] in Fig. 3A), we approximate $K$ for each ionic species to be a constant. We first evaluate $K$ of halide ions through the PS-SFVS measurement of $\sigma$ for a set of HX solutions mixed with NaX (again, X = Cl, Br, and I). With NaX added in a HX solution, $\rho(X^-)$ could be varied with [X$^-$], dictated by $K(X^-)$, but $\rho(Na^+)$ is negligible because Na$^+$ ion is known to be repelled by its own image from the water/vapor surface (3, 13, 18, 34), and $\rho(H_3O^+)$ follows [H$^+$] to be invariant. Therefore, we have $\sigma$ varying with [NaX] through $\rho(X^-)$, and can obtain $K(X^-)$ from $K(X^-) \approx (-1/|e|) \cdot \partial\sigma/\partial[X^-]$. Fig. 4 shows the measured $\sigma$ for the surface of HX–NaX mixture solutions as a function of the bulk halide concentrations with different [H$^+$]. It is seen that $\sigma$ increases (decreases) with an increase in [H$^+$] ([X$^-$]), as expected for larger amounts of the adsorbed $H_3O^+$ (X$^-$). As [Br$^-$] or [I$^-$] is large, $\sigma$ can eventually change its sign (becomes negative), i.e., $\rho(X^-) > \rho(H_3O^+)$. The observed $\sigma$ versus [X$^-$] for each halide species is fitted by lines with a constant slope, yielding $K(X^-)$= –0.54 (±0.62), 2.68 (±1.28), and 4.62 (±0.98) Å for Cl$^-$, Br$^-$, and I$^-$, respectively. Note that $K(I^-) > K(Br^-) > K(Cl^-)$ and $K(Cl^-) \approx \rho(Cl^-) \approx 0$, i.e., no preferential surface adsorption of Cl$^-$ ions, are in agreement with the MD simulations (3) and earlier qualitative SFVS studies (13, 34).

With $K(X^-)$ determined, we can further use it to estimate $\rho(X^-)$ from [X⁻] for a given pure HX solution, and then extract $\rho(H_3O^+)$ of such a surface from the difference between the measured $\sigma/|e|$ and the estimated $\rho(X^-)$. $\rho(H_3O^+)$ so obtained for HCl, HBr, and HI solutions are presented in Fig. 3B as a function of bulk ion concentrations (see *SI Appendix*, section 3 for a detailed analysis). It is found that $\rho(H_3O^+)$ versus [HX] is consistent for distinct halide species within our measurement uncertainty, even if the corresponding $\sigma/|e|$ appears with different values (Fig. 3A). The lack of specific anion effect on $\rho(H_3O^+)$ supports the argument that the correlation between ions caused by direct ion-ion interactions is not important for the concentrations studied. Therefore, our observation reveals the intrinsic characteristics of a hydronium ion at the water/vapor interface. A linear fit for $\rho(H_3O^+)$ versus [H⁺] in Fig. 3B unveils the partitioning coefficient $K(H_3O^+) = 11.2\ (\pm 1.0)$ Å.

One could convert the deduced $K$ to the partitioning coefficient defined as the ratio of the volume ion density in the surface region to that in the bulk, denoted by $K_V$, if the depth profile of ion distributions can be known. Because such information is not experimentally available, we follow the theoretical reports (4, 8, 36) to consider the adsorbed hydronium ions to distribute in a surface layer with an effective thickness $L$ that ranges from a H-bonding length (1.97 Å) to an intermolecular distance (~3.1 Å). This approximation yields $K_V \approx K/L = 3.61 \sim 5.68$, corresponding to a pH difference of the surface with respect to the bulk region, $\Delta$pH, by –0.66 ($\pm 0.10$), and the adsorption free energy for hydronium ions $\Delta G = -3.74\ (\pm 0.56)$ kJ/mol. Note that the deduced $K_V$, $\Delta$pH, and $\Delta G$, being intrinsic characteristics of hydronium, are applicable to aqueous interfaces with even lower ion concentrations, so we have the surface pH of 6.34 ($\pm 0.10$) at bulk pH 7.

A brief comparison between our result and the reported experiments is given for hydronium ions, but the discussion about the adsorption of halide ions will appear elsewhere. For surface tension measurements, the work by Pegram and Record based on a partitioning model provides an estimate of $K_V$ of ~1.5 for hydronium, corresponding to $\Delta G$ about –1.0 kJ/mol (18), but Levin *et. al.* yields a very different $\Delta G$ ($\approx$ –7.5 kJ/mol) by employing polarizable anion dielectric continuum theory (36). A recent analysis of the surface tension data with the help of classical MD simulations gives another value of $\Delta G \approx$ –2.5 kJ/mol (37). $\Delta G$ deduced from our spectroscopic work is within the range of these literature values, but, apparently, the estimates from the surface tension depend very much on the model adopted. In addition, Tahara *et. al.*, used the electronic SFG to probe acid-base equilibrium of pH indicator dyes on the water surface, suggesting a $\Delta$pH of $-1.7^{+1.0}_{-1.1}$ (16). While the average value is significantly higher than our result from the intrinsic water surface ($\Delta$pH = –0.66±0.10), the uncertainty of their result is too large to justify if it is affected by the binding interaction between protons and the dye molecule.

Note that we rely solely on the data with low molar fraction of ions (< 0.5 %, depicted as solid dots in Fig. 3) to derive the above conclusions, whereas analyses of the high-concentration data are difficult. First, both $\Delta\chi^{(2)}_{BIL}$ and $\chi^{(2)}_{EDL}$ may vary with the bulk ion concentration simultaneously as the surface ion density is large, as discussed earlier. Second, the GC theory adopted in our analysis approximates ions as point charges, but this assumption may break down by the steric effect of ions at high concentrations. Here this problem is excluded in our specific study by comparing the GC theory with a modified GC theory (38) that takes the finite-size effect of the ions into account (see *SI Appendix*, section 4 for a detailed analysis). Third, it is unclear whether $\chi^{(3)}(z) \cong \chi^{(3)}_B$ remains valid or not for an EDL with high molar fractions of ions (above a few %) (22, 23), or $E_0(z)$ that appears largely within the region of the BIL (Debye length is ~0.3 nm for [HX] = 1M), because of distortion of the H-bonding structure. We have performed a separate PS-SHG analysis of $\sigma$ for the HCl-solution/vapor interface. Results of $\sigma$ from PS-SHG and PS-SFVS are found consistent for [HCl] up to 0.8 M (see *SI Appendix*, section 5 for details), implying that the above three approximations are still valid. But, we cannot rule out the possibility of their failure when the ion concentration is high.

We have also attempted to quantify the hydroxide ions at the NaOH-solution/vapor interface using the same PS-SFVS scheme. The ion-induced spectral change is detectable only for [NaOH] > ~1 M (not shown here), which does not allow us to conduct reliable analysis with the assumptions justified.

**Conclusion**

We have conducted a PS-SFVS analysis to quantify the affinity of hydronium ions at the water/vapor interface. Adsorption of $H_3O^+$ is shown to follow a constant partitioning coefficient and independent of specific hydronium-anion interactions, and thus unveils the intrinsic characteristics of a hydronium ion at the water surface. The adsorption free energy is measured to be $\Delta G = -3.74\ (\pm 0.56)$ kJ/mol, corresponding to a pH difference of $-0.66\ (\pm 0.10)$ at the surface with respect to the bulk value. Our findings are not only of importance for better understanding of atmospheric processes, but also offer a microscopic-level basis to develop advanced quantum-mechanical models for molecular simulations of nuclear quantum effects in hydrogen-bonded systems, and sophisticated theories for interpreting macroscopic observables of aqueous interfaces, e. g., surface tension and electrokinetic zeta potential.

# Materials and Methods:

**Sample preparations**

HCl (37 wt %, reagent grade), NaCl (> 99 %, reagent), HBr (48 wt %, ACS reagent), NaBr (> 99 %, reagent), HI (55 wt %, ACS reagent), and NaI (> 99.5 %, ACS reagent) were purchased from Sigma-Aldrich. NaCl and NaBr were baked at 600 °C for 6 hours and slowly cooled down to room temperature before usage. The rest of chemicals were used as received. To prevent the photochemistry of iodide from occurring, the HI and HI–NaI mixture solutions were stored in dark environment and were prepared at most 2 hours in advance of their use. Ultra-pure distilled water was obtained from Milli-Q water purification system with the resistivity of 18.2 MΩ-cm. All the glassware was sonicated with diluted Deconex 11 Universal solution for 15 mins, and then rinsed thoroughly with ultra-pure water.

**Phase-sensitive sum-frequency vibrational spectroscopy (PS-SFVS)**

Our broadband PS-SFVS setup has been reported previously (24, 26). The light source was an optical parametric amplifier pumped by a 1-kHz femtosecond Ti:sapphire laser system (Astrella, Coherent), combined with a difference frequency generator and a grating-based spectral filter for generating tunable mid-IR pulses (with a bandwidth of ~250 cm$^{-1}$) and narrow-band visible pulses ($\omega_{vis}$ ~12500 cm$^{-1}$ with a bandwidth of ~13.2 cm$^{-1}$). The mid-IR and picosecond visible pulses were focused onto the sample surface, after co-propagating through a *y*-cut quartz acting as the local oscillator (LO) and a SrTiO$_3$ plate for phase modulation. The SFG spectral interferogram created by the sample in reflection and the LO reflected from the sample was measured by a charge-coupled device (CCD)-based polychromator, and Fourier-analyzed (29) to yield the complex

$\chi^{(2)}_{S,eff}(\omega_{IR})$ after normalization against a *z*-cut quartz crystal. The measurements were performed with S-, S-, P- (SSP-) polarized SFG, visible, and mid-IR fields at room temperature. The Fresnel coefficients were removed from the presented data.

**Acknowledgments:** We thank Y. R. Shen, Yuki Nagata, and Mischa Bonn for helpful discussions; L.D. acknowledges support from Academia Sinica. **Funding:** This work was funded by the Ministry of Science and Technology, Taiwan (grant number 106-2112-M-001-001-MY3; 108-2112-M-001-039-MY3). **Author contributions:** K.Y.C. and Y.C.W. designed the research project, and performed the experiments and analyses. K.Y.C., L.D., and Y.C.W. discussed the results and wrote the manuscript. **Competing interests:** The authors declare that they have no competing interest. **Data and materials availability:** All data needed to evaluate the conclusions in the paper are present in the paper and/or the Supplementary Information.

# References

1. (1987) *The Chemistry of Acid Rain: Sources and Atmospheric Processes* (American Chemical Society: Washington, DC).
2. Irwin JG & Williams ML (1988) Acid rain: Chemistry and transport. *Environmental Pollution* 50(1-2):29-59.
3. Jungwirth P & Tobias DJ (2006) Specific ion effects at the air/water interface. *Chemical Reviews* 106(4):1259-1281.
4. Pezzotti S & Gaigeot MP (2018) Spectroscopic BIL-SFG invariance hides the chaotropic effect of protons at the air-water interface. *Atmosphere* 9(10):396.
5. Buch V, Milet A, Vacha R, Jungwirth P, & Devlin JP (2007) Water surface is acidic. *Proceedings of the National Academy of Sciences of the United States of America* 104(18):7342-7347.
6. Dang LX (2003) Solvation of the hydronium ion at the water liquid/vapor interface. *Journal of Chemical Physics* 119(12):6351-6353.
7. Ishiyama T & Morita A (2007) Molecular dynamics analysis of interfacial structures and sum frequency generation spectra of aqueous hydrogen halide solutions. *Journal of Physical Chemistry A* 111(38):9277-9285.
8. Tse YLS, Chen C, Lindberg GE, Kumar R, & Voth GA (2015) Propensity of hydrated excess protons and hydroxide anions for the air-water interface. *Journal of the American Chemical Society* 137(39):12610-12616.
9. Ottosson N, *et al.* (2011) Increased propensity of I-aq(-) for the water surface in non-neutral solutions: Implications for the interfacial behavior of H3Oaq+ and OHaq. *Journal of Physical Chemistry Letters* 2(9):972-976.
10. Shapovalov VL, Mohwald H, Konovalov OV, & Knecht V (2013) Negligible water surface charge determined using Kelvin probe and total reflection X-ray fluorescence techniques. *Physical Chemistry Chemical Physics* 15(33):13991-13998.
11. Petersen PB & Saykally RJ (2005) Evidence for an enhanced hydronium concentration at the liquid water surface. *Journal of Physical Chemistry B* 109(16):7976-7980.
12. Tarbuck TL, Ota ST, & Richmond GL (2006) Spectroscopic studies of solvated hydrogen and hydroxide ions at aqueous surfaces. *Journal of the American Chemical Society* 128(45):14519-14527.
13. Gopalakrishnan S, Liu DF, Allen HC, Kuo M, & Shultz MJ (2006) Vibrational spectroscopic studies of aqueous interfaces: Salts, acids, bases, and nanodrops. *Chemical Reviews* 106(4):1155-1175.
14. Tian CS, Ji N, Waychunas GA, & Shen YR (2008) Interfacial structures of acidic and basic aqueous solutions. *Journal of the American Chemical Society* 130(39):13033-13039.
15. Hua W, Verreault D, & Allen HC (2015) Relative Order of Sulfuric Acid, Bisulfate, Hydronium, and Cations at the Air-Water Interface. *Journal of the American Chemical Society* 137(43):13920-13926.
16. Yamaguchi S, Kundu A, Sen P, & Tahara T (2012) Communication: Quantitative estimate of the water surface pH using heterodyne-detected electronic sum frequency generation. *Journal of Chemical Physics* 137(15):151101.
17. Agmon N, *et al.* (2016) Protons and Hydroxide Ions in Aqueous Systems. *Chemical Reviews* 116(13):7642-7672.

18. Pegram LM & Record MT (2006) Partitioning of atmospherically relevant ions between bulk water and the water/vapor interface. *Proceedings of the National Academy of Sciences of the United States of America* 103(39):14278-14281.
19. Dreier LB, Bemhard C, Gonella G, Backus EHG, & Bonn M (2018) Surface potential of a planar charged lipid-water interface. what do vibrating plate methods, second harmonic and sum frequency measure? *Journal of Physical Chemistry Letters* 9(19):5685-5691.
20. Ong SW, Zhao XL, & Eisenthal KB (1992) Polarization of water molecules at a charged interface: second harmonic studies of the silica/water interface *Chemical Physics Letters* 191(3-4):327-335.
21. Dreier LB*, et al.* (2018) Saturation of charge-induced water alignment at model membrane surfaces. *Science Advances* 4(3):eaap7415.
22. Pezzotti S, Galimberti DR, Shen YR, & Gaigeot MP (2018) Structural definition of the BIL and DL: a new universal methodology to rationalize non-linear chi((2))(omega) SFG signals at charged interfaces, including chi((3))(omega) contributions. *Physical Chemistry Chemical Physics* 20(7):5190-5199.
23. Joutsuka T & Morita A (2018) Electrolyte and temperature effects on third-order susceptibility in sum-frequency generation spectroscopy of aqueous salt solutions. *Journal of Physical Chemistry C* 122(21):11407-11413.
24. Wen Y-C*, et al.* (2016) Unveiling microscopic structures of charged water interfaces by surface-specific vibrational spectroscopy. *Physical Review Letters* 116:016101.
25. Gonella G, Lutgebaucks C, de Beer AGF, & Roke S (2016) Second harmonic and sum-frequency generation from aqueous interfaces is modulated by interference. *Journal of Physical Chemistry C* 120(17):9165-9173.
26. Wen YC, Zha S, Tian CS, & Shen YR (2016) Surface pH and ion affinity at the alcohol-monolayer/water interface studied by sum-frequency spectroscopy. *Journal of Physical Chemistry C* 120(28):15224-15229.
27. Ohno PE*, et al.* (2019) Beyond the Gouy-Chapman Model with Heterodyne-Detected Second Harmonic Generation. *Journal of Physical Chemistry Letters* 10(10):2328-2334.
28. Ohno PE, Wang HF, & Geiger FM (2017) Second-order spectral lineshapes from charged interfaces. *Nature Communications* 8:1032.
29. Mondal JA, Nihonyanagi S, Yamaguchi S, & Tahara T (2012) Three distinct water structures at a zwitterionic lipid/water interface revealed by heterodyne-detected vibrational sum frequency generation. *Journal of the American Chemical Society* 134(18):7842-7850.
30. Ji N, Ostroverkhov V, Tian CS, & Shen YR (2008) Characterization of vibrational resonances of water-vapor interfaces by phase-sensitive sum-frequency spectroscopy. *Physical Review Letters* 100(9):096102.
31. Bard AJ & Faulkner LR (1980) *Electrochemical Methods: Fundamentals and Applications* (Wiley, New York).
32. Stiopkin IV*, et al.* (2011) Hydrogen bonding at the water surface revealed by isotopic dilution spectroscopy. *Nature* 474(7350):192-195.
33. The phase factor $\exp(i\Delta k_z z)$ in Eq. (1) has negligible influences on interpretation of the EDL spectra in Fig. 2 [phase of $\exp(i\Delta k_z z)$ is < 7º for bulk ion concentrations > 3 mM].
34. Tian C, Byrnes SJ, Han H-L, & Shen YR (2011) Surface propensities of atmospherically relevant ions in salt solutions revealed by phase-sensitive sum frequency vibrational spectroscopy. *Journal of Physical Chemistry Letters* 2(15):1946-1949.

35. The average separation between ions in this study is larger than the Bjerrum length in water, suggesting that correlation between ions caused by their electrostatic interactions is not important.
36. dos Santos AP & Levin Y (2010) Surface tensions and surface potentials of acid solutions. *Journal of Chemical Physics* 133(15):154107.
37. Mamatkulov SI, Allolio C, Netz RR, & Bonthuis DJ (2017) Orientation-induced adsorption of hydrated protons at the air-water Interface. *Angewandte Chemie-International Edition* 56(50):15846-15851.
38. Borukhov I, Andelman D, & Orland H (1997) Steric effects in electrolytes: A modified Poisson-Boltzmann equation. *Physical Review Letters* 79(3):435-438.

# Figure Captions:

**Fig. 1.** $\text{Im}\chi^{(2)}_{S,eff}(\omega_{IR})$ SFG spectra of the water/vapor interface. (*A*) Spectra in the complete OH-stretch frequency range for different HCl concentrations in water. (*B*) and (*C*) Spectra with different hydrogen halides in water (~0.1 M in *B* and 0.8 M in *C*). Dots and black solid lines in *A*, *B*, and *C* denote the results of hydrogen halide solutions and the neat water, respectively. Red line in *C* is a guide to the eyes. $\text{Im}\chi^{(2)}_{S,eff}(\omega_{IR})$ has an unit of $10^{-21}$ m²/V.

**Fig. 2.** Spectra of complex $\chi^{(2)}_{EDL}$ of electrical double layer for different HCl concentrations in water. (*A*) and (*B*) Imaginary and real parts of $\chi^{(2)}_{EDL}$ spectra with the unit of $10^{-22}$ m²/V, respectively. Dots are the measured results, and lines are fitting curves based on the GC theory and the $\chi^{(3)}_B(\omega_{IR})$ spectrum of bulk water.

**Fig. 3.** Characterization of charges/ions at the water/vapor interface. (*A*) Surface charge density deduced from PS-SFVS for different hydrogen halide concentrations in water. (*B*) Surface density of hydronium ions, $\rho(H_3O^+)$, obtained from the data in *A*. In *A* and *B*, solid dots are results with the approximations used in the analysis fully justified, whereas the open dots (for ~0.8 M HCl, ~0.8 M HBr, and >0.3 M HI) may have assumption-related concerns (see text for details). Lines are fitting curves assuming a constant ratio between $\sigma$ (or $\rho$) and the bulk HX concentration.

**Fig. 4.** Surface charge density $\sigma$ deduced from PS-SFVS for surfaces of the HX–NaX mixture solutions [X = I, Br, and Cl, for (*A*), (*B*), and (*C*), respectively]. Blue, black, and red dots are results with [H⁺] = 10, 33, and ~100 mM, respectively, controlled by [HX] in water. Open and solid dots

denote data obtained from pure HX solutions and the HX–NaX mixtures, respectively. Lines are fitting curves assuming a constant ratio between $\sigma$ and [X$^-$] for each halide species.

**Figures:**

# Figure 1

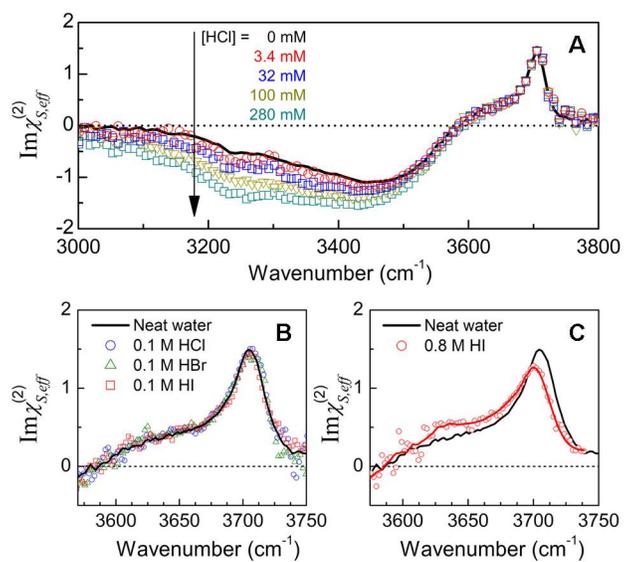

# Figure 2

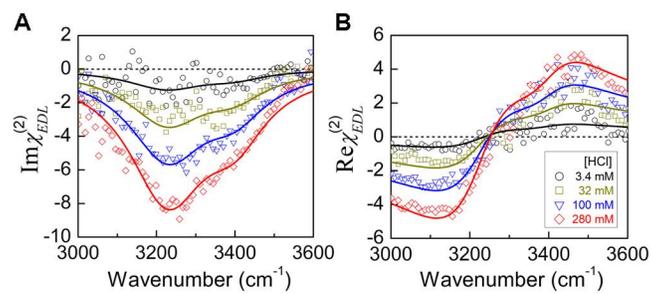

# Figure 3

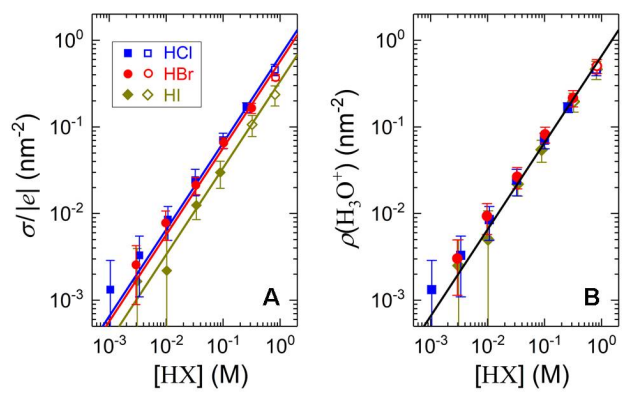

**Figure 4**

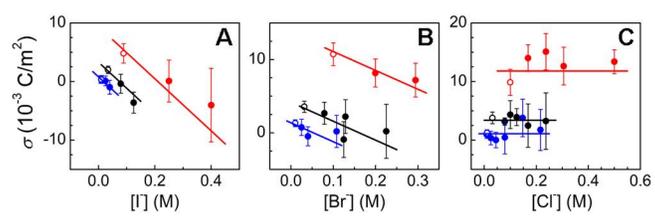